\documentstyle[12pt,fullpage,subeqn]{article}

\def\bib{\bibitem}
\def\be{\begin{equation}}
\def\ee{\end{equation}}
\def\beq{\begin{eqnarray}}
\def\eeq{\end{eqnarray}}
\def\barr{\begin{array}}
\def\earr{\end{array}}

\def\etal{{\it et al.}}
\def\lsim{\:\raisebox{-0.5ex}{$\stackrel{\textstyle<}{\sim}$}\:}
\def\gsim{\:\raisebox{-0.5ex}{$\stackrel{\textstyle>}{\sim}$}\:}
\def\rp{$R_p \hspace{-1em}/\;\:$}

\def\gev{\; \rm  GeV}

\def\nun{m_{\tilde \nu_{nL}} }
\def\utn{m_{\tilde u_{nL}} }
\def\dln{m_{\tilde d_{nL}} }
\def\drn{m_{\tilde d_{nR}} }
\def\dis{\displaystyle}
\def\ib#1,#2,#3{           {\it ibid.\/ }{\bf #1} (19#2) #3}
\def\ap#1,#2,#3{           {\it Ann. Phys. (NY)\/ }{\bf #1} (19#2) #3}
\def\ijmp#1,#2,#3{         {\it Int. J. Mod. Phys.\/ } {\bf A#1} (19#2) #3}
\def\mpl#1,#2,#3 {          {\it Mod. Phys. Lett.\/ } {\bf A#1} (19#2) #3}
\def\np#1,#2,#3{           {\it Nucl. Phys.\/ }{\bf B#1} (19#2) #3}
\def\npps#1,#2,#3{         {\it Nucl. Phys. B (Proc. Suppl.)\/ }{\bf B#1}
                             (19#2) #3}
\def\plb#1,#2,#3{           {\it Phys. Lett.\/ }{\bf B#1} (19#2) #3}
\def\pr#1,#2,#3{           {\it Phys. Rev.\/ }{\bf D#1} (19#2) #3}
\def\prep#1,#2,#3{         {\it Phys. Rep.\/ }{\bf #1} (19#2) #3}
\def\prl#1,#2,#3{          {\it Phys. Rev. Lett.\/ }{\bf #1} (19#2) #3}
\def\pro#1,#2,#3{          {\it Prog. Theor. Phys.\/ }{\bf #1} (19#2) #3}
\def\rmp#1,#2,#3{          {\it Rev. Mod. Phys.\/ }{\bf #1} (19#2) #3}
\def\sjnp#1,#2,#3{         {\it Sov. J. Nucl. Phys.\/ }{\bf #1} (19#2) #3}
\def\sp#1,#2,#3{           {\it Sov. Phys.-Usp.\/ }{\bf #1} (19#2) #3}
\def\zp#1,#2,#3{           {\it Zeit. f\"ur Physik\/ }{\bf #1} (19#2) #3}
\begin{document}
\setcounter{page}{0}
\renewcommand{\thefootnote}{\fnsymbol{footnote}}
\thispagestyle{empty}
\vspace*{-1in}
\begin{flushright}
MPI--PTh/96--20\\
TIFR/TH/96--12\\[2ex]
{\large \tt hep-ph/9603363} \\
\end{flushright}
\vskip 45pt
\begin{center}
{\Large{\bf New Constraints On Lepton Nonconserving R-parity
Violating Couplings}} \\[2cm]
{\large Debajyoti Choudhury}\footnote{debchou@mppmu.mpg.de}\\
\vspace{10pt}
{\em Max--Planck--Institut f\"ur Physik, Werner--Heisenberg--Institut,\\
F\"ohringer Ring 6, 80805 M\"unchen,  Germany.}

\vspace{12pt}
\rm and

\vspace{12pt}

{\large Probir Roy}\footnote{probir@theory.tifr.res.in} \\
\vspace{10pt}
{\em Institute of Theoretical Physics, Santa Barbara, CA93106, U.S.A} \\
and \\
{\em Theoretical Physics Group, Tata Institute of Fundamental Research, \\
    Homi Bhabha Road,  Bombay 400005,
India}\footnote{Permanent address}

\vspace{50pt}
{\bf ABSTRACT}
\end{center}

\begin{quotation}

Strong upper bounds are derived on certain product combinations of
lepton nonconserving couplings in the minimal supersymmetric standard
model with explicit $R$-parity violation.  The input is information
from rare leptonic decays of the long-lived neutral kaon, the muon and
the tau as well as from the mixings of neutral $K$- and $B$-mesons.
One of these bounds is comparable and another superior to
corresponding ones obtained recently from neutrinoless double beta
decay.

\end{quotation}

\vfill
\newpage
\setcounter{footnote}{0}
\renewcommand{\thefootnote}{\arabic{footnote}}

\setcounter{page}{1}
\pagestyle{plain}
\advance \parskip by 10pt

The minimal supersymmetric standard model MSSM~\cite{mssm}
is now a leading
candidate for physics beyond the standard model.  A natural question
to ask is whether global conservation laws (such as those of
baryon number $B$ and lepton number $L$), valid in the standard model,
hold as one
goes to MSSM.  This is related to the question of $R$-parity
($R_p = (-1)^{3B+L+2 S}$, with $S$ the intrinsic spin~\cite{rpardef})
conservation for which no credible theoretical argument is in
existence.  It is therefore important to consider the phenomenology of
possible $R_p$-violating terms~\cite{rpar,rpar2} in the Lagrangian.
We direct our attention to scenarios~\cite{rpar,rpar2}
of explicit $R_p$-violation instead of
those with a spontaneous one since the former can be accommodated
within the particle/sparticle spectrum of MSSM while the latter needs
at least one additional singlet superfield.

It is difficult to
accommodate both $L$-violating and $B$-violating
 \rp\ terms within the constraints of proton
decay~\cite{vissani}.
Assuming that the baryon-number
violating terms are identically zero~\cite{b-cons} helps
evade this constraint in a natural way, apart from rendering
simpler the cosmological
requirement of the survival of GUT baryogenesis through to the
present day.  This survival can then be assured if
at least one of the lepton numbers $L_i$ is conserved over
cosmological time scales~\cite{cosmo}.

The $R_p$-violating and
lepton-nonconserving part of the superpotential, with the MSSM
superfields in usual notation, involves two kinds of couplings
$\lambda_{ijk}$ and $\lambda'_{ijk}$ ($i,j,k$ being family indices),
and can be written as
\be
W = \frac{1}{2} \lambda_{ijk} L_i L_j \overline{E_k}
  + \lambda'_{ijk} L_i Q_j \overline{D_k},
     \label{superpot}
\ee
where $\lambda_{ijk} = -\lambda_{jik}$. We have here omitted possible
bilinear terms as they are not relevant to the discussion.
Most phenomenological studies of these couplings
have been aimed at deriving upper bounds~\cite{bgh}
 on the magnitudes of
individual $\lambda$- or $\lambda'$-coupling in terms of observed or
unobserved processes.  However, recently there has been some interest
in deriving bounds on the products~\cite{crs,vissani,bm,agashe}
 of two such couplings which
are stronger than the products of known bounds on the respective
individual couplings.  This will be the approach taken here.

Before such an analysis is attempted, one needs to consider the
possible effects of fermion mixing on the
\rp\ sector~\cite{vissani,agashe}, if only because this can itself
be a cause for the simultaneous presence of more than one such coupling.
These effects are twofold. For one, it is, in some sense, more natural
to consider eq.(\ref{superpot}) to be defined in the {\em gauge basis}
rather than in the {\em mass basis}. In that case, even if there were
only one particular nonzero $\lambda'$--coupling, quark mixing would
generate a plethora of such couplings, albeit related to one another.
In this Letter,
we do not confine ourselves
to this particular theoretical motivation, but rather consider
the most general lepton number violating \rp\ sector.
The second effect is more subtle. Since the Cabibbo-Kobayashi-Maskawa
matrix $K$
is different from identity, the $SU(2)_L$ symmetry of the $\lambda'$
terms is no longer manifest when rexpressed in terms
of the mass eigenstates. To wit,
\be
W_{\lambda'} = \lambda'_{ijk} \left( N_i D_j - \sum_p K^{\dag}_{jp} E_i U_p
                              \right) \overline{D_k},
     \label{lam-pr}
\ee
where the $\lambda'_{ijk}$ have already been redefined to absorb some
field rotation effects. In eq.(\ref{lam-pr}) we have chosen to suppress
the effects of possible non-alignment of fermion and sfermion mass matrices.
However, since any such non-alignment is already constrained
severely~\cite{susy_const}, it is reasonable to neglect such interplay of
different effects.

 In terms of the component fields, the  interaction
term in the Lagrangian density can then be expressed as
\be
\barr{rcl}
{\cal L}_I & = & \dis
   \lambda_{ijk}
         \left(  \tilde\nu_{iL} \overline{e_{kR}} e_{jL}
               + \tilde e_{jL} \overline{e_{kR}} \nu_{iL}
               + {\tilde e}^\star_{kR} \overline{(\nu_{iL})^C} e_{jL}
         \right)
                     \\[2ex]
& + & \dis
  \lambda'_{ijk}
      \left[  \tilde \nu_{iL} \overline{d_{kR}} d_{jL}
            + \tilde d_{jL} \overline{d_{kR}} \nu_{iL}
            + {\tilde d}_{kR}^\star \overline{(\nu_{iL})^C} d_{jL}
      \right.  \\[1.5ex]
&& \dis \hspace*{2em}
      \left.
          - \sum_p K^\dagger_{jp}
            \left(  \tilde e_{iL} \overline{d_{kR}} u_{pL}
                  + \tilde u_{pL} \overline{d_{kR}} e_{iL}
                  + {\tilde d}^\star_{kR} \overline{(e_{iL})^C} u_{pL}
            \right)
      \right]
       + h.c.
\earr
        \label{lagrangian}
\ee
Here $e (\tilde e)$ stands for a charged lepton (slepton) and the
fields for the other particles have been designated by the
corresponding name letters.

\begin{table}[h]
\begin{center}
\begin{tabular}{|l|l||l|l|}
\hline
\multicolumn{1}{|c|}{Quantity} & \multicolumn{1}{c||}{Experimental} &
\multicolumn{1}{|c|}{Quantity} & \multicolumn{1}{c|}{Experimental} \\
& \multicolumn{1}{|c||}{value (bound)} & 
    & \multicolumn{1}{|c|}{value (bound)}  \\
&&&\\[-1.5ex]
\hline
&&& \\[-1.5ex]
$ \delta m_K $ & $3.5 \times 10^{-12}$ MeV   
    &   $ \delta m_B $ & $3.4 \times 10^{-10}$ MeV   \\
&&&\\[-1.5ex]
\hline
&&&\\[-1.5ex]
$ {\rm Br }(K_L \rightarrow \bar e e) $ & $ < 4.1 \times 10^{-11} $   &   
$ {\rm Br }(K_S \rightarrow \bar e e) $ & $ < 1.0 \times 10^{-5} $   \\
$ {\rm Br }(K_L \rightarrow \bar \mu \mu) $ & 
               $ ( 7.4 \pm 0.4) \times 10^{-11} $   &   
$ {\rm Br }(K_S \rightarrow \bar \mu \mu) $ & 
               $ < 3.2 \times 10^{-7} $   \\   
$ {\rm Br }(K_L \rightarrow \bar \mu e + \mu \bar e ) $ & 
               $ < 3.3 \times 10^{-11} $   &   
$ {\rm Br }(K^+ \rightarrow \pi \nu \bar\nu) $ & 
               $ < 5.2 \times 10^{-9} $   \\ 
&&&\\[-1.5ex]
\hline
&&&\\[-1.5ex]
$ {\rm Br }(\mu \rightarrow 3 e) $ & $ < 10^{-12} $ &
$ {\rm Br }(\tau \rightarrow 3 e) $ & $ < 1.3 \times 10^{-5} $ \\
$ {\rm Br }(\tau \rightarrow \bar \mu e e ) $ & $ < 1.4 \times 10^{-5} $ &
$ {\rm Br }(\tau \rightarrow \bar e e \mu) $ & $ < 1.4 \times 10^{-5} $ \\
$ {\rm Br }(\tau \rightarrow \bar \mu \mu e ) $ & $ < 1.9 \times 10^{-5} $ &
$ {\rm Br }(\tau \rightarrow \bar e \mu \mu) $ & $ < 1.6 \times 10^{-5} $ \\
$ {\rm Br }(\tau \rightarrow 3 \mu) $ & $ < 1.7 \times 10^{-5} $ & & \\[1.5ex]
\hline
\end{tabular}
\end{center} 
\caption{{\em Experimental numbers~\protect\cite{pdg}
 used as upper bounds on the contribution of the \rp\ terms
to the various observables.} }
         \label{exptal}
\end{table}

We first consider the constraints on products of $\lambda'$-couplings
given by our knowledge of the neutral $B$- and $K$-meson mass levels.
The mass difference $\delta m_B$ between $B_1$ and $B_2$ (and
similarly $\delta m_K$ between $K_1$ and $K_2$) arises out of the
mixing between the $B^0_d$, $\overline{B^0_d}$ (and $K^0$,
$\overline{K^0}$) mesons.  The $\lambda'$-couplings of
eq.(\ref{lagrangian}) make such mixing
amplitudes possible at the tree-level through the exchange
of a sneutrino $\tilde\nu_i$ both in the $s$- and $t$-channels.  In
fact, the
effective Lagrangian terms for $B_d - \overline{B_d}$ and
$K - \overline{K}$ mixings are
\subequations
\be \displaystyle
{\cal L}_{eff} (B \rightarrow \overline{B})
  = - \sum_n \frac{\dis \lambda'_{n31} \lambda^{\prime\star}_{n13} }
                          {m^2_{\tilde\nu_n}} \;
   \overline{d_R} b_L \ \overline{d_L} b_R,
       \label{bb:effec}
\ee
\be \displaystyle
{\cal L}_{eff} (K \rightarrow \overline{K})
  =  - \sum_n \frac{\dis \lambda'_{n21} \lambda^{\prime\star}_{n12} }
                          {m^2_{\tilde\nu_n}} \;
   \overline{d_R} s_L \ \overline{d_L} s_R,
       \label{kk:effec}
\ee
\endsubequations

We calculate the contributions of (\ref{bb:effec}) and
(\ref{kk:effec}) to $\delta m_B$,
$\delta m_K$ respectively and require them not to exceed the
corresponding experimental numbers (see Table~\ref{exptal}).
Though it is tempting to improve the bounds by first
subtracting the SM contributions, we choose not do so as the latter
involve considerable uncertainties, both experimental and theoretical.
Before we present our numbers, it is useful (for notational purposes)
to define the following ``normalized'' quantities~:
\be
\barr{rclcrcl}
n_n & \equiv & \dis \left( \frac{100 \gev} {\nun} \right)^2
& \qquad &
u_n & \equiv & \dis \left( \frac{100 \gev} {\utn} \right)^2 \\[2ex]
d^L_n & \equiv & \dis \left( \frac{100 \gev} {\dln} \right)^2
& \qquad &
d^R_n & \equiv & \dis \left( \frac{100 \gev} {\drn } \right)^2 .
\earr
           \label{defn:mass}
\ee

Finally, our numbers from $\delta m_B$ and
$\delta m_K$ respectively are\footnote{Here, and in the rest of the
discussion, bounds on any (complex) quantity will apply to its magnitude.}:
\subequations
\be
\sum_i \lambda_{i31}^{\prime \ast} \lambda'_{i13} n_i
          \lsim 3.3 \times 10^{-8},
    \label{bb:constr}
\ee
\be
\sum_i \lambda_{i21}^{\prime \ast} \lambda'_{i12} n_i
              \lsim 4.5 \times 10^{-9}.
    \label{kk:constr}
\ee
\endsubequations
The upper bound in eq.(\ref{bb:constr})
is only marginally weaker than that ($3 \times 10^{-8}$)
  obtained by Babu and Mohapatra~\cite{bm}
 very recently from the
lack of observation of neutrinoless double-beta decay.  In contrast,
the bound of eq.(\ref{kk:constr})
is three orders of magnitude stronger than the
$10^{-6}$ obtained by those authors.

\begin{table}[h]
\begin{center}
\begin{tabular}{|c|c|c|}
\hline
Decay Mode & Combinations constrained & Upper bound \\
     &  &  \\[-1.5ex]
\hline
&& \\[-1.5ex]
&  $\lambda_{122} \lambda'_{112} n_1, \ 
    \lambda_{122} \lambda'_{121} n_1, \
    \lambda_{232} \lambda'_{312} n_3, \
    \lambda_{232} \lambda'_{321} n_3$ 
&  $  3.8 \times 10^{-7}$ \\[0.75ex]
\cline{2-3}
&&\\[-4ex]
$K_L \rightarrow \mu \bar \mu$ & \\[-1ex]
& ($\star$) \hspace{1em} 
  $\lambda^{\prime}_{211} \lambda'_{222} (u_1 + u_2), \
   \lambda^{\prime}_{212} \lambda'_{221} (u_1 + u_2)$
& $ 3.3 \times 10^{-5}$ \\
&& \\[-1.5ex]
\hline
&& \\[-1.5ex]
$K_L \rightarrow e\bar e$ 
&  $ \lambda_{121} \lambda'_{212} n_2, \ 
     \lambda_{121} \lambda'_{221} n_2, \
     \lambda_{131} \lambda'_{312} n_3, \
     \lambda_{131} \lambda'_{321} n_3$ 
& $ 2.5 \times 10^{-8}$ \\
&& \\[-1.5ex]
\hline
&& \\[-1.5ex]
& $\lambda_{122} \lambda'_{212} n_2, \
   \lambda_{122} \lambda'_{221} n_2, \
   \lambda_{132} \lambda'_{312} n_3, \
   \lambda_{132} \lambda'_{321} n_3$ & \\[-1ex]
& & $ 2.3 \times 10^{-8}$ \\[-1ex]
& $\lambda_{121} \lambda'_{112} n_1, \
   \lambda_{121} \lambda'_{121} n_1, \
   \lambda_{231} \lambda'_{312} n_3, \
   \lambda_{231} \lambda'_{321} n_3$ & \\[0.75ex]
  \cline{2-3}
&&\\[-4ex]
$K_L \rightarrow e\bar \mu + \bar e\mu$ & & \\[-1ex]
& $\lambda^{\prime}_{111} \lambda'_{212} u_1, \ 
   \lambda^{\prime}_{112} \lambda'_{211} u_2, \
   \lambda^{\prime}_{121} \lambda'_{222} u_2 $ & \\[-1ex]
&&   $ 3.5 \times 10^{-7}$ \\[-1ex]
& $\lambda^{\prime}_{122} \lambda'_{221} u_2, \
   \lambda^{\prime}_{131} \lambda'_{232} u_3, \
   \lambda^{\prime}_{132} \lambda'_{231} u_3$ & \\
&& \\[-1.5ex]
\hline
&& \\[-1.5ex]
$K^+ \rightarrow \pi \nu \bar\nu $ & 
$\lambda'_{i 1 n} \lambda'_{j 2 n} d^R_n, \ 
    \lambda'_{i n 2} \lambda'_{j n 1} d^L_n $ 
    \hspace{2em} ( For all $i,j, n$ )
& $ 4.8 \times 10^{-5}$ \\[1.5ex]
\hline
\end{tabular}
\end{center}
\caption{{\em Upper bounds on the magnitudes of coupling products derived 
from rare $K$ decays,
under the assumption that only one such 
product is nonzero. In the entry marked with ($\star$), we 
have assumed that $K_{12} = - K_{21}$.
For definitions, see eq.(\protect\ref{defn:mass}).} }
         \label{table:K}
\end{table}

Next, we focus on rare leptonic decay modes of the neutral
$K$-mesons: $K_{L,S} \rightarrow e \bar e$ or $\mu \bar
\mu$, $e\bar \mu$, $\mu\bar e$ as well as the semileptonic decay
$K^+ \rightarrow \pi^+ e_i \bar e_j$.  At the partonic level,
the generic subprocess is one in
which a down-type quark-antiquark pair ($d_k$ and $\bar d_\ell$, say)
tranform into a charged lepton-antilepton pair (assumed to be $e_i$ and
$\bar e_j$): $d_k + \overline{d_\ell} \rightarrow e_i + \bar e_j$.  Here
$i,j,k,\ell$ are generation indices.  The reaction can proceed via the
exchange of a $u$-squark $\tilde u_{nL}$ in the
$t$-channel as well
as via an $s$-channel exchange of a sneutrino $\tilde\nu_n$.  The
effective Lagrangian terms can be obtained in the same manner as above.
We have then
\be
\barr{rcl}
{\cal L}_{eff} (d_k + \bar d_\ell \rightarrow e_i + \bar e_j)
  & = & \dis
   \sum_n \left[  \frac{\lambda^{\star}_{nij} \lambda'_{nk\ell} }
                       {m^2_{\tilde\nu_n}}
                     \ \overline{d_{\ell R}} d_{kL}
                     \ \overline{e_{iL}} e_{jR}
                + \frac{\lambda^{\star}_{nji} \lambda'_{nlk} }
                       {m^2_{\tilde\nu_n}}
                     \ \overline{d_{\ell L}} d_{kR}
                     \ \overline{e_{iR}} e_{jL} \right]
         \\[2.8ex]
   &- & \dis
    \sum_{n,p} \frac{ K_{np} \lambda^{\prime\star}_{ipk} \lambda'_{jnl} }
                    { 2 \utn^2 } \
                 \overline{e_{iL}} \gamma_\mu e_{jL} \
                 \overline{d_{lR}} \gamma^\mu d_{kR} \ .
\earr
            \label{dd_ee}
\ee
The last term on the RHS comes from the $t$-channel exchange of a
$u$-squark while
the first two arise from the two $s$-channel diagrams (the first term
with the
sneutrino entering the hadronic vertex and the second with the
sneutrino leaving it). On calculating the relevant matrix element of the
effective Hamiltonian from (\ref{dd_ee}), we find that the new
contributions to the
decay $K_L \rightarrow e_i \bar e_j$ can be parametrized in terms of
the combinations
\be
\barr{rcl}
{\cal A}_{ij} & \equiv & \dis \sum_{n,p} u_n K_{np}
      \left(  \lambda^{\prime \star}_{ip1} \lambda^{\prime}_{jn2}
            - \lambda^{\prime \star}_{ip2} \lambda^{\prime}_{jn1}
      \right), \\[2ex]
{\cal B}_{ij} & \equiv & \dis \sum_n n_n \lambda^\star_{n i j}
      \left(  \lambda^{\prime}_{n12}- \lambda^{\prime}_{n21} \right).
\earr
      \label{K_L:combi}
\ee
Two points need to be emphasised here :
\begin{itemize}
  \item On account of the particular Lorentz structure of the operators,
        the contributions of the ${\cal A}_{ij}$ terms are
        chirality-suppressed. Hence, in the event of vanishing ${\cal
        B}_{ij}$, the
        constraints on ${\cal A}_{ij}$ would be weaker than those operative
        in the reverse case.
  \item In the limit of a trivial CKM matrix ($K_{np} = \delta_{np}$),
         ${\cal A}_{ij}^\ast = - {\cal A}_{ji}$, and thus
        ${\cal A}_{11}$ and ${\cal A}_{22}$ are purely imaginary.
        Though large imaginary parts in $\lambda'_{ijk}$  cannot be
        ruled out {\em per se}, these are liable to generate
        unacceptably large CP
        violating effects.
\end{itemize}
Nonetheless, we retain, for the time being,
the possibility of complex \rp\ couplings.
Ignoring again
the SM contributions, we demand that the \rp\ contribution
by itself does not exceed the experimental upper bounds. Considering
{}~\cite{pdg}
$K_{L} \rightarrow e \bar e$, $\mu \bar \mu$ and
$e\bar \mu + \mu\bar e$ in turn, we have
\be
\barr{rcl}
2.5 \times 10^{-8} & \gsim & \dis |{\cal B}_{11}| \\[1.5ex]
1.3 \times 10^{-13} & \gsim & \dis
    \left[ \left|{\cal B}_{22} \right|^2
          -  0.099 \; {\rm Re} \left({\cal B}_{22}^2  \right) \right]
    + 0.0024\;  \left| {\cal A}_{22} \right|^2
    + 0.10\; {\rm Im}( {\cal A}_{22}) \; {\rm Im} ({\cal B}_{22} )
              \\[1.5ex]
5.4 \times 10^{-16} & \gsim & \dis
    \left(  \left| {\cal B}_{12} \right|^2
          + \left| {\cal B}_{21} \right|^2 \right)
    + 0.0022 \; \left(  \left| {\cal A}_{12} \right|^2
                    + \left| {\cal A}_{21} \right|^2 \right) \\[1.2ex]
&& \dis \hspace*{5em}
    + 0.047 \; {\rm Re}\left( \{ {\cal A}_{12}^\ast - {\cal A}_{21} \}
                     {\cal B}_{12} \right)
\earr
\ee

The \rp\ contribution to the rare $K_S$ decays, on the other hand, are
parametrized by the combinations
\be
\barr{rcl}
{\cal C}_{ij} & \equiv & \dis \sum_{n,p} u_n K_{np}
      \left(  \lambda^{\prime \star}_{in1} \lambda^{\prime}_{jn2}
            + \lambda^{\prime \star}_{in2} \lambda^{\prime}_{jn1}
      \right), \\[2ex]
{\cal D}_{ij} & \equiv & \dis \sum_n n_n \lambda^\star_{n i j}
      \left(  \lambda^{\prime}_{n12} + \lambda^{\prime}_{n21} \right).
\earr
      \label{K_S:combi}
\ee
Since all of the relevant ${\cal C}_{ij}$s are real even in the limit of
a trivial CKM matrix, we cannot use
the arguments proffered in the case of the $K_L$ to disentangle the
contributions (the mass suppression continues nonetheless).
The resulting expressions are quite analogous to those for $K_L$ decays
and, in the case of ~\cite{pdg} $K_S \rightarrow \mu^- \mu^+$, looks like
\be
   \left[  \left|{\cal D}_{22} \right|^2
         + 0.099 \;{\rm Re} \left({\cal D}_{22}^2  \right) \right]
   + 0.1 \; {\rm Re}\left({\cal C}_{22}^\star {\cal D}_{22} \right)
   + 0.0025\; |{\cal C}_{22}|^2
  \lsim 3.1 \times 10^{-9}.
\ee

The additional contribution to the decay
$K^+ \rightarrow \pi^+ \nu_i \bar{\nu}_j$ due to the \rp\ terms
is in the form of two sets
of $t$--channel diagrams (for each $i,j$), one each with
$\tilde d_{Ln}$ and $\tilde d_{Rn}$. The effective Lagrangian can be
parametrized as
\be \dis
{\cal L}_{eff} (K^+ \rightarrow \pi^+ \nu_i \bar \nu_j)  =
     \frac{1}{2}\  \overline{\nu_{iL}} \gamma_\mu \nu_{jL} \
   \sum_n
     \left[  \frac{ \lambda^{\prime\star}_{i2n} \lambda'_{j1n} }
                {\drn^2 }
                \  \overline{s_L} \gamma^\mu d_L
         - \frac{\lambda^{\prime\star}_{in1} \lambda'_{jn2} }
                {\dln^2}
                \ \overline{s_R} \gamma^\mu d_R
\right].
            \label{k-pinunu:effec}
\ee
Since the nine possible combinations ($ij$) cannot be
distinguished experimentally, the bound ~\cite{pdg} from the non-observation
of such a mode leads to an upper bound on the incoherent
sum (and hence also on the individual quantities):
\be
         \sum_{i,j} \left| {\cal E}_{ij} \right|^2
           \lsim 2.3 \times 10^{-9},
            \label{k-pinunu:bound}
\ee
where
\be
{\cal E}_{ij}  \equiv  \dis \sum_n
      \left( d^R_n  \lambda^{\prime \star}_{i2n} \lambda^{\prime}_{j1n}
            - d^L_n \lambda^{\prime \star}_{in1} \lambda^{\prime}_{jn2}
      \right).
         \label{K+:combi}
\ee

\begin{table}[h]
\begin{center}
\begin{tabular}{||c|c|c||}
\hline
&& \\[0.5ex]
Decay Mode & Combinations constrained & Upper bound \\
     &  &  \\[-1.5ex]
\hline
&& \\[-1.5ex]
$\mu \rightarrow 3e$ 
& $ \lambda_{121} \lambda_{122} n_2, \ 
    \lambda_{131} \lambda_{132} n_3, \
    \lambda_{231} \lambda_{131} n_3$ 
& $6.6 \times 10^{-7}$ \\
& & \\[-1.5ex]
\hline
& & \\[-1.5ex]
$\tau^- \rightarrow 3e$ 
& $ \lambda_{121} \lambda_{123} n_2, \ 
    \lambda_{131} \lambda_{133} n_3, \ 
    \lambda_{231} \lambda_{121} n_2$
& $5.6 \times 10^{-3}$ \\  
& & \\[-1.5ex]
\hline
& & \\[-1.5ex]
$\tau^- \rightarrow e^- e^- \mu^+$ 
& $ \lambda_{231} \lambda_{112} n_2, 
    \lambda_{231} \lambda_{133} n_3$ 
& $5.7 \times 10^{-3}$ \\
& & \\[-1.5ex]
\hline
& & \\[-1.5ex]
$\tau^- \rightarrow \mu^- e^+ e^-$ 
& $ \lambda_{131} \lambda_{121} n_1, \ 
    \lambda_{122} \lambda_{123} n_2, \ 
    \lambda_{132} \lambda_{133} n_3, $ 
& $5.7 \times 10^{-3}$ \\ 
& $ \lambda_{232} \lambda_{121} n_2, \ 
    \lambda_{131} \lambda_{233} n_3$ & \\
& & \\[-1.5ex]
\hline
& & \\[-1.5ex]
$\tau^- \rightarrow e^+ \mu^- \mu^-$ 
& $\lambda_{131} \lambda_{121}  n_1, \ 
   \lambda_{132} \lambda_{233}  n_3$ 
& $6.2 \times 10^{-3}$ 
\\
& & \\[-1.5ex]
\hline
& & \\[-1.5ex]
$\tau^- \rightarrow e^- \mu^+ \mu^-$ 
& $ \lambda_{131} \lambda_{122} n_1, \ 
    \lambda_{232} \lambda_{133} n_3, \ 
    \lambda_{232} \lambda_{122} n_2,$ 
& $6.7 \times 10^{-3}$ \\
& $ \lambda_{121} \lambda_{123}  n_1, \
    \lambda_{231} \lambda_{233} n_3$ & \\
& & \\[-1.5ex]
\hline
& & \\[-1.5ex]
$\tau \rightarrow 3\mu$ 
&  $ \lambda_{132} \lambda_{122} n_1, \ 
     \lambda_{122} \lambda_{123} n_1, \ 
     \lambda_{232} \lambda_{233} n_3$
& $6.4 \times 10^{-3}$ \\[1.5ex]
\hline
\end{tabular}
\end{center}
\caption{{\em Upper bounds on the magnitudes of 
coupling products derived from flavour
changing $lepton$ decays, under the assumption that only one such 
product is nonzero. For definitions, 
see eq.(\protect\ref{defn:mass}).} }
        \label{table:L}
\end{table}

Finally, we come to products of $\lambda$-couplings which are probed
by information on rare three-body leptonic decays of $\mu$ and $\tau$.
All these
processes can be characterized by the generic transition
$e_a \rightarrow e_b + e_c + \bar e_d$
where $a,b,c,d$ are generation indices.  The
reaction proceeds by the exchange of a sneutrino $\tilde\nu_n$
in the $t$-channel as well as in the $u$-channel. Analogous to the
neutral $K$-decays, here too the sneutrino propagator may have
both directions. The effective Lagrangian term now is
\be
\barr{rcl}
\dis (100 \gev)^2 \;
{\cal L}_{eff} (e_a \rightarrow e_b + e_c + \bar e_d) & = & \dis
     {\cal F}_{abcd}\; \overline{e_{bR}} e_{aL}
                     \ \overline{e_{cL}} e_{dR}
   + {\cal F}_{dcba}\; \overline{e_{bL}} e_{aR}
                     \ \overline{e_{cR}} e_{dL}    \\[2ex]
 &+& {\cal F}_{acbd}\; \overline{e_{cR}} e_{aL}
                     \ \overline{e_{bL}} e_{dR}
   + {\cal F}_{dbca}\; \overline{e_{cL}} e_{aR}
                     \ \overline{e_{bR}} e_{dL},
\earr
        \label{l_dk:eff}
\ee
where
\be  \dis
     {\cal F}_{abcd} \equiv
       \sum_n n_n \lambda_{nab} \lambda^\star_{ncd}\ .
\ee
In eq.(\ref{l_dk:eff}) the first (last) two terms on the RHS
correspond to the $t$- ($u$-) channel exchange diagrams.
Utilizing the known experimental
upper limits ~\cite{pdg} on the relevant partial widths, we have,
\be
\barr{lcl}
\dis \mu \rightarrow 3 e & : \qquad & \dis
           \left| {\cal F}_{1112} \right|^2
         + \left| {\cal F}_{2111} \right|^2 \lsim 4.3 \times 10^{-13},
    \\[2ex]
\dis \tau \rightarrow 3 e & : \qquad & \dis
           \left| {\cal F}_{1113} \right|^2
         + \left| {\cal F}_{3111} \right|^2 \lsim 3.1 \times 10^{-5},
    \\[2ex]
\dis \tau \rightarrow 3 \mu & : \qquad & \dis
           \left| {\cal F}_{2223} \right|^2
         + \left| {\cal F}_{3222} \right|^2 \lsim 4.1 \times 10^{-5},
    \\[2ex]
\dis \tau^- \rightarrow \mu^+ e^- e^- & : \qquad & \dis
           \left| {\cal F}_{3112} \right|^2
         + \left| {\cal F}_{2113} \right|^2 \lsim 3.3 \times 10^{-5},
    \\[2ex]
\dis \tau^- \rightarrow e^+ \mu^- \mu^- & : \qquad & \dis
           \left| {\cal F}_{3221} \right|^2
         + \left| {\cal F}_{1223} \right|^2 \lsim 3.8 \times 10^{-5},
    \\[2ex]
\dis \tau^- \rightarrow e^+ e^- \mu^- & : \qquad & \dis
           \left| {\cal F}_{1123} \right|^2
         + \left| {\cal F}_{3211} \right|^2
         + \left| {\cal F}_{1213} \right|^2
         + \left| {\cal F}_{3121} \right|^2 \lsim 3.3 \times 10^{-5},
    \\[2ex]
\dis \tau^- \rightarrow e^- \mu^+ \mu^- & : \qquad & \dis
           \left| {\cal F}_{3122} \right|^2
         + \left| {\cal F}_{2213} \right|^2
         + \left| {\cal F}_{3212} \right|^2
         + \left| {\cal F}_{2123} \right|^2 \lsim 4.5 \times 10^{-5}.
\earr
     \label{l_dk:res}
\ee

It is thus clear that the rare decays considered here lead to
myriad bounds on combinations of \rp\ couplings.
Since particular products do occur more than once (albeit with
different scalar masses), it is instructive to ask what the bounds
would be if {\em only one} product were non-zero.
In Table~\ref{table:K},
we give the constraints, under such assumptions, for
the products relevant to $K$-decays.  The corresponding $\lambda \lambda$
products (which are constrained from flavour violating lepton
decays) are listed in Table~\ref{table:L}.
We should mention here that weaker constraints on
$n_2 |\lambda_{121} \lambda_{122}|$ and
$n_3 |\lambda_{131} \lambda_{132}|$ were earlier estimated from the
non-observation of $\mu \rightarrow 3e$ decay~\cite{hinchliffe}.

To conclude, we have derived quite strong upper bounds on certain
product combinations of $\lambda$ and $\lambda'$ couplings.  The
reason that the bounds are so restrictive is that they come from
processes which are permitted by such couplings at the tree-level but
are disallowed or have to proceed via loops
both in the SM and in the MSSM.  If we consider transitions
such as $\mu \rightarrow e\gamma$ or $\tau \rightarrow e\gamma$, which
are loop-induced even with such couplings, the
corresponding bounds would not be anything like as strong.  Of course,
our list is not fully exhaustive -- all possible coupling
product combinations have not been covered.
It is interesting to note that, barring the constraints from
$\mu \rightarrow 3 e$, the bounds on the $\lambda \lambda$ products
are, in general, weaker than those on the $\lambda \lambda'$
combinations\footnote{Since $\lambda$ and $\lambda'$ couplings mimic
a class of dilepton and leptoquark phenomenology respectively, it is likely
that a similar conclusion may be reached in the generic case as
well~\protect\cite{frank}.}.
On the other hand, every single
$\lambda$-coupling appears in Tables 1 and 2.  The same is not the
case for the
$\lambda'$-couplings.  For instance, $\lambda'_{322}$
and $\lambda'_{323}$ are
unconstrained while couplings like $\lambda'_{22k}, \lambda'_{13k}$
(except $\lambda'_{133}$) are only weakly constrained~\cite{bgh}.  Thus
one may still harbor good hope of detecting a nonzero signal, say at LEP2,
from some of these couplings~\cite{lep2}.
A large top quark event sample, accumulated at Fermilab, will also
provide \cite{dreiner-phillips} a good opportunity to
probe certain \rp\ couplings from possible bounds on
nonstandard top decays.

This research was initiated during a visit to CERN by P.R. who
acknowledges the hospitality of the Theory Division.  It has been
supported in part by the National Science Foundation under Grant No.
PHY94-07194.  We thank
G. Bhattacharyya, H.~Dreiner and H.~Murayama for their helpful remarks.

\end{document}